\documentclass[prc,preprint,eqsecnum,nofootinbib,amsmath,amssymb,preprintnumbers,
tightenlines,longbibliography]{revtex4}

\usepackage{afterpage}
\usepackage{mathrsfs}
\usepackage{graphicx}% Include figure files
\usepackage{bm}% bold math
\usepackage{paralist}
\usepackage[colorlinks,linkcolor=blue,citecolor=blue]{hyperref}

\advance\parskip 1.9pt
\advance\voffset -0.2in

\begin{document}

\title{Multi-particle long-range rapidity correlations from 
fluctuation of the fireball longitudinal shape
} 

\author{Adam~Bzdak}
\email{bzdak@fis.agh.edu.pl}
\affiliation
    {%
     AGH University of Science and Technology, Faculty of Physics and
     Applied Computer Science, al. Mickiewicza 30, 30-059 Krakow, Poland
    }%

\author{Piotr Bo\.zek}
\email{piotr.bozek@fis.agh.edu.pl}
\affiliation
    {%
     AGH University of Science and Technology, Faculty of Physics and
     Applied Computer Science, al. Mickiewicza 30, 30-059 Krakow, Poland
    }%

\begin{abstract}
We calculate the genuine long-range multi-particle rapidity correlation
functions, $C_{n}(y_1,\dots,y_n)$ for $n=3,4,5,6$, originating from
fluctuations of the fireball longitudinal shape. In these correlation
functions any contribution from the short-range two-particle correlations, and 
in general up to $(n-1)$-particle in $C_n$, is
suppressed. The information about the fluctuating fireball shape in rapidity
is encoded in the cumulants of coefficients of the orthogonal polynomial expansion of
particle distributions in rapidity.
\end{abstract}
%\date{\today}
\maketitle

\section{Introduction}

Fluctuations in the longitudinal structure of the fireball produced in
heavy-ion collisions has drawn noticeable interest in recent years, see,
e.g., \cite%
{ATLAS:2015kla,Bialas:2011bz,Bzdak:2012tp,Bozek:2010vz,Bozek:2015bha,Bialas:2011xk, Olszewski:2015xba,Cheng:2011hz,Csernai:2012mh,Pang:2012he,Vovchenko:2013viu,Jia:2014ysa,Pang:2014pxa}. 
These fluctuations result in new phenomena and modify known correlations
in rapidity and azimuthal angle.

In Ref. \cite{Bzdak:2012tp} it was argued that fluctuations of the fireball
longitudinal shape result in the specific long-range two-particle rapidity
correlations that depend not only on the rapidity difference, $y_{1}-y_{2}$,
but also on the rapidity sum, $y_{1}+y_{2}$. In analogy to the long-range
azimuthal correlations originating from fluctuating shape of the fireball in
the transverse direction \cite{Voloshin:2008dg,Alver:2010gr}, it was
proposed to expand the single particle rapidity distribution in terms of the
orthogonal polynomials 
\begin{equation}
\rho (y;a_{0},a_{1},...)=\rho (y)\left[ 1+\sum\nolimits_{i=0}a_{i}T_{i}%
\left( y\right) \right] ,  \label{ro-an}
\end{equation}%
where $\rho (y)\equiv \langle \rho (y)\rangle $ is the measured single
particle distribution. $a_{0}$ represents rapidity independent multiplicity
fluctuation of the fireball as a whole, $a_{1}$ is an event-by-event
asymmetric component\footnote{%
In the wounded nucleon model \cite{Bialas:1976ed,Bialas:2004su} and for
symmetric A+A collisions, $a_{1}$ corresponds to the difference between
left- and right-going wounded nucleons, $a_{1}\propto w_{L}-w_{R}$.}, etc.,
see Ref. \cite{Bzdak:2012tp} for more details. Averaging Eq. (\ref{ro-an})
over $a_{0},a_{1},...$ with the probability distribution $P(a_{0},a_{1},...)$
we obtain $\langle a_{i}\rangle =0$.

The two-particle rapidity distribution at a given $a_{0},a_{1},...$ is%
\begin{equation}
\rho _{2}(y_{1},y_{2};a_{0},a_{1},...)=\rho (y_{1};a_{0},a_{1},...)\rho
(y_{2};a_{0},a_{1},...)+\mbox{short\ range}\,,
\end{equation}%
where we added the short-range correlations in rapidity, that cannot be
written, on an event-by-event basis, as a product of fluctuating
single-particle distributions (\ref{ro-an}). If resonance decays or jets are
the dominant mechanism behind the short-range contribution, we expect this
correlation to be approximately a function of $y_{1}-y_{2}$ with a range of
about one unit in rapidity.

Taking an average over $a_{i}$ and subtracting $\rho (y_{1})\rho (y_{2})$,
we obtain the final two-particle rapidity correlation function%
\begin{eqnarray}
C_{2}(y_{1},y_{2}) &=&\rho _{2}(y_{1},y_{2})-\rho (y_{1})\rho (y_{2})  \notag
\\
&=&\rho (y_{1})\rho (y_{2})\left[ \sum\nolimits_{i,k=0}\left( \langle
a_{i}a_{k}\rangle +b_{ik}\right) T_{i}\left( y_{1}\right) T_{k}\left(
y_{2}\right) \right] ,  \label{C2}
\end{eqnarray}%
where $b_{ik}$ denotes some short-range (SR) correlations not related to the
fluctuating shape of the fireball 
\begin{equation}
b_{ik}=\int_{-1}^{1}{dy_{1}}{dy_{2}}\frac{C_{2}^{\text{SR}%
}(y_{1},y_{2})T_{i}(y_{1})T_{k}(y_{2})}{\rho (y_{1})\rho (y_{2})},
\label{srcor}
\end{equation}%
where $C_{2}^{\text{SR}}(y_{1},y_{2})$ is the short-range two-particle
correlation function. It is essential to remove this unwanted background and
this is the subject of the paper.

We propose to measure the cumulants of the genuine multi-particle rapidity
correlation functions in analogy to the multi-particle flow cumulants \cite%
{Borghini:2000sa,Borghini:2001vi}, which proved to be effective in removing
non-flow effects from correlations in azimuthal angle \cite%
{Khachatryan:2015waa}.

Throughout the paper we use $y=\frac{\eta }{Y}$, where $\eta $ is rapidity
or pseudorapidity in the range $[-Y,Y]$. The orthogonalization condition for
the Legendre polynomials\footnote{%
In Ref. \cite{Bzdak:2012tp} we expanded the distribution (\ref{ro-an}) in
the Chebyshev polynomials but other choices are certainly possible.} \cite%
{Jia:2015jga} reads $\int {dy}T_{i}(y)T_{k}(y)=\delta _{ik}$, and $T_{k}(y)=%
\sqrt{k+\frac{1}{2}}P_{k}\left( \frac{\eta }{Y}\right) $ with $P_{k}(x)=%
\frac{1}{2^{k}k!}\frac{d^{k}}{dx^{k}}(x^{2}-1)^{k}$.

In the next Section we derive formulas for the genuine three-, four-, five-
and six-particle correlation functions originating from fluctuating
longitudinal shape of the fireball. We discuss our results in Section 3.

\section{Multi-particle correlations}

In this section we discuss multi-particle correlations originating from
fluctuating fireball shape in rapidity. At this point it is useful to
comment on the experimental way of estimating the integrals over the
multi-particle distributions. In the experimental analysis \cite%
{ATLAS:2015kla}, estimates of $\langle a_{i}a_{k}\rangle $ are obtained from
the integration of the two-particle correlation function. It is challenging
to apply this method to the higher order cumulants. However, the standard
procedure of summing over n-tuples, e.g., 
\begin{eqnarray}
\langle T_{i}(y_{a})T_{j}(y_{b})T_{k}(y_{c})T_{l}(y_{d})\rangle  &\equiv
&\int_{-1}^{1}{dy_{1}}{dy_{2}}{dy_{3}}{dy_{4}}\frac{\rho
_{4}(y_{1},y_{2},y_{3},y_{4})T_{i}(y_{1})T_{j}(y_{2})T_{k}(y_{3})T_{l}(y_{4})%
}{\rho (y_{1})\rho (y_{2})\rho (y_{3})\rho (y_{4})}  \notag \\
&=&\left\langle {\sum_{a,b,c,d}}^{^{\prime }}\frac{T_{i}(y_{a})}{\rho (y_{a})%
}\frac{T_{j}(y_{b})}{\rho (y_{b})}\frac{T_{k}(y_{c})}{\rho (y_{c})}\frac{%
T_{l}(y_{d})}{\rho (y_{d})}\right\rangle ,  \label{T-exp}
\end{eqnarray}%
can be applied for samples with sufficient statistics \cite{Bilandzic:2010jr}%
. $\rho _{4}(y_{1},y_{2},y_{3},y_{4})$ is the measured four-particle
rapidity density. In the last line of the above expression the sum runs over
four different particles in a given event and the average is over all events.%
\footnote{%
By definition $\langle T_{i}(y_{a})\rangle =\int {dy}T_{i}(y)\sim \delta
_{i,0}$.}

\subsection{Three-particle correlations}

The three-particle distribution at a given $a_{0},a_{1},...$ is%
\begin{equation}
\rho _{3}(y_{1},y_{2},y_{3};a_{0},a_{1},...)=\rho
(y_{1};a_{0},a_{1},...)\rho (y_{2};a_{0},a_{1},...)\,\rho
(y_{3};a_{0},a_{1},...)+\mbox{short\ range}\,.
\end{equation}%
Expanding on the orthogonal basis and taking an average over $a_{i}$ we
obtain%
\begin{eqnarray}
\frac{\rho _{3}(y_{1},y_{2},y_{3})}{\rho (y_{1})\rho (y_{2})\rho (y_{3})}
&=&1+\sum\nolimits_{i,k=0}\left( \left\langle a_{i}a_{k}\right\rangle
+b_{ik}\right) \left[ T_{i}\left( y_{1}\right) T_{k}\left( y_{2}\right)
+T_{i}\left( y_{1}\right) T_{k}\left( y_{3}\right) +T_{i}\left( y_{2}\right)
T_{k}\left( y_{3}\right) \right] +  \notag \\
&&\sum\nolimits_{i,k,m=0}\left( \left\langle a_{i}a_{k}a_{m}\right\rangle
+b_{ikm}\right) T_{i}\left( y_{1}\right) T_{k}\left( y_{2}\right)
T_{m}\left( y_{3}\right) \,,
\end{eqnarray}%
where $b_{ik}$ and $b_{ikm}$ represent the unwanted short-range background
(not related to fluctuations of the fireball longitudinal structure). We are
interested in extracting information about the genuine three-particle
correlations\footnote{%
In general, the genuine $n$-particle correlation function $%
C_{n}(y_{1},...,y_{n})$ is non-zero only if there is a physical mechanism
directly correlating $n$ or more particles.}, $C_{3}(y_{1},y_{2},y_{3})$,
defined as%
\begin{eqnarray}
\rho _{3}(y_{1},y_{2},y_{3}) &=&\rho (y_{1})\rho (y_{2})\rho (y_{3})+\rho
(y_{1})C_{2}(y_{2},y_{3})+\rho (y_{2})C_{2}(y_{1},y_{3})+  \notag \\
&&\rho (y_{3})C_{2}(y_{1},y_{2})+C_{3}(y_{1},y_{2},y_{3}),  \label{rho-3}
\end{eqnarray}%
where $\rho _{3}(y_{1},y_{2},y_{3})$ is the three-particle rapidity density.
If genuine three-particle short-range correlations can be neglected, $%
b_{ikm}=0$, performing simple calculations we obtain 
\begin{equation}
C_{3}(y_{1},y_{2},y_{3})=\rho (y_{1})\rho (y_{2})\rho (y_{3})\left[
\sum\nolimits_{i,k,m=0}\left\langle a_{i}a_{k}a_{m}\right\rangle T_{i}\left(
y_{1}\right) T_{k}\left( y_{2}\right) T_{m}\left( y_{3}\right) \right] .\,
\label{C3}
\end{equation}

Equation (\ref{C3}) allows to directly extract $\left\langle
a_{i}a_{k}a_{m}\right\rangle $ without the contribution from the
two-particle short-range correlations $b_{ik}$ 
\begin{equation}
\left\langle a_{i}a_{k}a_{m}\right\rangle _{[3]}=\int dy_{1}dy_{2}dy_{3}%
\frac{C_{3}(y_{1},y_{2},y_{3})T_{i}(y_{1})T_{k}(y_{2})T_{m}(y_{3})}{\rho
(y_{1})\rho (y_{2})\rho (y_{3})},
\end{equation}%
where $\left\langle ...\right\rangle _{[3]}$ denotes\footnote{%
For $C_{3}$ we have $\langle a_{i}a_{k}a_{m}\rangle _{\lbrack 3]}=\langle
a_{i}a_{k}a_{m}\rangle $, which is not the case for higher order correlation
functions.} that $\langle a_{i}a_{k}a_{m}\rangle $ is sensitive to $C_{3}$
but it does not depend on the lower order correlation function $C_{2}$.

Using Eq. (\ref{rho-3}) we can relate $\langle a_{i}a_{k}a_{m}\rangle
_{\lbrack 3]}$ through integrals of the two- and three-particle densities $%
\rho _{n}$ and finally through sums over n-tuples, see Eq. (\ref{T-exp}) 
\begin{eqnarray}
\left\langle a_{i}a_{k}a_{m}\right\rangle _{[3]} &=&\left\langle
T_{i}(y_{a})T_{k}(y_{b})T_{m}(y_{c})\right\rangle -\left\langle
T_{i}(y_{a})\right\rangle \left\langle T_{k}(y_{a})T_{m}(y_{b})\right\rangle
-\left\langle T_{k}(y_{a})\right\rangle \left\langle
T_{i}(y_{a})T_{m}(y_{b})\right\rangle -  \notag \\
&&\left\langle T_{m}(y_{a})\right\rangle \left\langle
T_{i}(y_{a})T_{k}(y_{b})\right\rangle +2\left\langle
T_{i}(y_{a})\right\rangle \left\langle T_{k}(y_{a})\right\rangle
\left\langle T_{m}(y_{a})\right\rangle   \notag \\
&\equiv &\left\langle T_{i}(y_{a})T_{k}(y_{b})T_{m}(y_{c})\right\rangle
_{[3]}.
\end{eqnarray}

Perhaps $C_{3}$ is not particularly useful because $\langle a_{i}^{3}\rangle 
$ is expected to be rather small if not zero (by definition $\langle
a_{i}\rangle =0$). Specific effects of correlations resulting from bias of
event multiplicity on rapidity distribution, e.g., stronger longitudinal
expansion in events with higher fireball density, can be tested using mixed
correlations 
\begin{equation}
\langle a_{0}a_{k}^{2}\rangle _{\lbrack 3]}=\left\langle
T_{0}(y_{a})T_{k}(y_{b})T_{k}(y_{c})\right\rangle -\left\langle
T_{0}(y_{a})\right\rangle \left\langle T_{k}(y_{a})T_{k}(y_{b})\right\rangle
,
\end{equation}%
where $\langle T_{0}(y_{a})\rangle =\sqrt{2}$ for the chosen normalization
and $k>0$.

\subsection{Four-particle correlation}

Here we discuss the more interesting case of four-particle correlation
function. Performing analogous calculations we obtain%
\begin{eqnarray}
\frac{\rho _{4}(y_{1},y_{2},y_{3},y_{4})}{\rho (y_{1})\rho (y_{2})\rho
(y_{3})\rho (y_{4})} &=&1+\sum\nolimits_{i,k=0}\left\langle
a_{i}a_{k}\right\rangle [T_{i}\left( y_{1}\right) T_{k}\left( y_{2}\right)
+T_{i}\left( y_{1}\right) T_{k}\left( y_{3}\right) +T_{i}\left( y_{1}\right)
T_{k}\left( y_{4}\right) +  \notag \\
&&T_{i}\left( y_{2}\right) T_{k}\left( y_{3}\right) +T_{i}\left(
y_{2}\right) T_{k}\left( y_{4}\right) +T_{i}\left( y_{3}\right) T_{k}\left(
y_{4}\right) ]+  \notag \\
&&\sum\nolimits_{i,k,m=0}\left\langle a_{i}a_{k}a_{m}\right\rangle
[T_{i}\left( y_{1}\right) T_{k}\left( y_{2}\right) T_{m}\left( y_{3}\right)
+T_{i}\left( y_{1}\right) T_{k}\left( y_{2}\right) T_{m}\left( y_{4}\right) +
\notag \\
&&T_{i}\left( y_{1}\right) T_{k}\left( y_{3}\right) T_{m}\left( y_{4}\right)
+T_{i}\left( y_{2}\right) T_{k}\left( y_{3}\right) T_{m}\left( y_{4}\right)
]+  \notag \\
&&\sum\nolimits_{i,k,m,n=0}\left\langle a_{i}a_{k}a_{m}a_{n}\right\rangle
T_{i}\left( y_{1}\right) T_{k}\left( y_{2}\right) T_{m}\left( y_{3}\right)
T_{n}\left( y_{4}\right) \,.
\end{eqnarray}

For simplicity we omit the short-range background correlations $b_{ik}$, $%
b_{ikm}$ and $b_{ikmn}$. We are interested in the genuine four-particle
correlation function, $C_{4}(y_{1},y_{2},y_{3},y_{4})$, defined as%
\begin{eqnarray}
\rho _{4}(y_{1},y_{2},y_{3},y_{4}) &=&\rho (y_{1})\rho (y_{2})\rho
(y_{3})\rho (y_{4})+\rho (y_{1})\rho (y_{2})C_{2}(y_{3},y_{4})+\rho
(y_{1})\rho (y_{3})C_{2}(y_{2},y_{4})+  \notag \\
&&\rho (y_{1})\rho (y_{4})C_{2}(y_{2},y_{3})+\rho (y_{2})\rho
(y_{3})C_{2}(y_{1},y_{4})+\rho (y_{2})\rho (y_{4})C_{2}(y_{1},y_{3})+  \notag
\\
&&\rho (y_{3})\rho (y_{4})C_{2}(y_{1},y_{2})+\rho
(y_{1})C_{3}(y_{2},y_{3},y_{4})+\rho (y_{2})C_{3}(y_{1},y_{3},y_{4})+  \notag
\\
&&\rho (y_{3})C_{3}(y_{1},y_{2},y_{4})+\rho
(y_{4})C_{3}(y_{1},y_{2},y_{3})+C_{2}(y_{1},y_{2})C_{2}(y_{3},y_{4})+  \notag
\\
&&C_{2}(y_{1},y_{3})C_{2}(y_{2},y_{4})+C_{2}(y_{1},y_{4})C_{2}(y_{2},y_{3})+C_{4}(y_{1},y_{2},y_{3},y_{4}).
\end{eqnarray}

Performing straightforward calculations we obtain%
\begin{equation}
\frac{C_{4}(y_{1},y_{2},y_{3},y_{4})}{\rho (y_{1})\rho (y_{2})\rho
(y_{3})\rho (y_{4})}=\sum\nolimits_{i,k,m,n=0}\left\langle
a_{i}a_{k}a_{m}a_{n}\right\rangle _{[4]}T_{i}\left( y_{1}\right) T_{k}\left(
y_{2}\right) T_{m}\left( y_{3}\right) T_{n}\left( y_{4}\right) ,  \label{C4}
\end{equation}%
where%
\begin{equation}
\left\langle a_{i}a_{k}a_{m}a_{n}\right\rangle _{[4]}\equiv \left\langle
a_{i}a_{k}a_{m}a_{n}\right\rangle -\left\langle a_{i}a_{k}\right\rangle
\left\langle a_{m}a_{n}\right\rangle -\left\langle a_{i}a_{m}\right\rangle
\left\langle a_{k}a_{n}\right\rangle -\left\langle a_{i}a_{n}\right\rangle
\left\langle a_{k}a_{m}\right\rangle .
\end{equation}%
with $\left\langle ...\right\rangle _{[4]}$ denoting that the object depends
on $C_{4}$ but not on the lower order correlations.

At this stage we are mostly interested in extracting the leading terms, $%
i=k=m=n$ (and in particular the asymmetric term $a_{1}$)%
\begin{eqnarray}
\langle a_{i}^{4}\rangle _{\lbrack 4]} &\equiv &\langle a_{i}^{4}\rangle
-3\langle a_{i}^{2}\rangle ^{2}  \notag \\
&=&\int dy_{1}dy_{2}dy_{3}dy_{4}\frac{C_{4}(y_{1},y_{2},y_{3},y_{4})T_{i}%
\left( y_{1}\right) T_{i}\left( y_{2}\right) T_{i}\left( y_{3}\right)
T_{i}\left( y_{4}\right) }{\rho (y_{1})\rho (y_{2})\rho (y_{3})\rho (y_{4})}
\notag \\
&=&\langle T_{i}(y_{a})T_{i}(y_{b})T_{i}(y_{c})T_{i}(y_{d})\rangle -3\langle
T_{i}(y_{a})T_{i}(y_{b})\rangle ^{2}\,,  \label{a1-cum}
\end{eqnarray}%
where in the last line of the above equation ($i>0$) we show the way to
calculate the cumulant, see Eq. (\ref{T-exp}).

Another interesting term in the expansion (\ref{C4}) is the mixed term $%
\langle a_{0}^{2}a_{k}^{2}\rangle $ for $k>0$. The expression for such a
cumulant reads 
\begin{eqnarray}
\langle a_{0}^{2}a_{k}^{2}\rangle _{\lbrack 4]} &\equiv &\langle
a_{0}^{2}a_{k}^{2}\rangle -\left\langle a_{0}^{2}\right\rangle \langle
a_{k}^{2}\rangle -2\langle a_{0}a_{k}\rangle ^{2}  \notag \\
&=&\langle T_{0}(y_{a})T_{0}(y_{b})T_{k}(y_{c})T_{k}(y_{d})\rangle
-\left\langle T_{0}(y_{a})T_{0}(y_{b})\rangle \right\langle
T_{k}(y_{a})T_{k}(y_{b})\rangle -  \notag \\
&&2\langle T_{0}(y_{a})T_{k}(y_{b})\rangle ^{2}-2\langle T_{0}(y_{a})\rangle
\langle T_{0}(y_{a})T_{k}(y_{b})T_{k}(y_{c})\rangle +  \notag \\
&&2\langle T_{0}(y_{a})\rangle ^{2}\langle T_{k}(y_{a})T_{k}(y_{b})\rangle .
\end{eqnarray}%
This expression removes two and three-particle correlations and correctly
takes into account that $\langle T_{0}(y_{a})\rangle \neq 0$. In particular $%
\langle a_{0}^{2}a_{2}^{2}\rangle _{\lbrack 4]}$ could be a measure of the
genuine correlations between event multiplicity and the width of the
particle distribution in rapidity.

\subsection{Five-particle correlation}

For the genuine five-particle correlation function, $C_{5}(y_{1},...,y_{5})$%
, defined as\footnote{%
For clarity we skip the argument $y_{i}$ and show only the numbers of
possible combinations.}%
\begin{equation}
\rho _{5}=\rho \rho \rho \rho \rho +\underset{5}{\underbrace{\rho C_{4}}}+%
\underset{10}{\underbrace{\rho \rho C_{3}}}+\underset{10}{\underbrace{\rho
\rho \rho C_{2}}}+\underset{15}{\underbrace{\rho C_{2}C_{2}}}+\underset{10}{%
\underbrace{C_{2}C_{3}}}+C_{5},
\end{equation}%
we obtain 
\begin{equation}
\frac{C_{5}(y_{1},...,y_{5})}{\rho (y_{1})...\rho (y_{5})}%
=\sum\nolimits_{i,k,m,n,r=0}\left\langle
a_{i}a_{k}a_{m}a_{n}a_{r}\right\rangle _{[5]}T_{i}\left( y_{1}\right)
T_{k}\left( y_{2}\right) T_{m}\left( y_{3}\right) T_{n}\left( y_{4}\right)
T_{r}\left( y_{5}\right) ,  \label{C5}
\end{equation}%
where%
\begin{equation}
\langle a_{i}^{5}\rangle _{\lbrack 5]}\equiv \left\langle
a_{i}a_{k}a_{m}a_{n}a_{r}\right\rangle -[\underset{10\text{ variations}}{%
\underbrace{\left\langle a_{i}a_{k}\right\rangle \left\langle
a_{m}a_{n}a_{r}\right\rangle +...}}].
\end{equation}

The leading term is%
\begin{eqnarray}
\langle a_{i}^{5}\rangle _{\lbrack 5]} &\equiv &\langle a_{i}^{5}\rangle
-10\langle a_{i}^{2}\rangle \langle a_{i}^{3}\rangle \\
&=&\langle
T_{i}(y_{a})T_{i}(y_{b})T_{i}(y_{c})T_{i}(y_{d})T_{i}(y_{e})\rangle
-10\langle T_{i}(y_{a})T_{i}(y_{b})\rangle \langle
T_{i}(y_{a})T_{i}(y_{b})T_{i}(y_{c})\rangle ,  \notag
\end{eqnarray}%
where in the second line of the above equation we assume $i>0$.

\subsection{Six-particle correlation}

Finally, for the six-particle correlation function, $C_{6}(y_{1},...,y_{6})$%
, defined as%
\begin{eqnarray}
\rho _{6} &=&\rho \rho \rho \rho \rho \rho +\underset{6}{\underbrace{\rho
C_{5}}}+\underset{15}{\underbrace{\rho \rho C_{4}}}+\underset{20}{%
\underbrace{\rho \rho \rho C_{3}}}+\underset{15}{\underbrace{\rho \rho \rho
\rho C_{2}}}+\underset{60}{\underbrace{\rho C_{2}C_{3}}}+\underset{45}{%
\underbrace{\rho \rho C_{2}C_{2}}}+\underset{15}{\underbrace{C_{2}C_{4}}}+ 
\notag \\
&&\underset{10}{\underbrace{C_{3}C_{3}}}+\underset{15}{\underbrace{%
C_{2}C_{2}C_{2}}}+C_{6},
\end{eqnarray}%
we obtain%
\begin{equation}
\frac{C_{6}(y_{1},...,y_{6})}{\rho (y_{1})...\rho (y_{6})}%
=\sum\nolimits_{i,k,m,n,r,s=0}\left\langle
a_{i}a_{k}a_{m}a_{n}a_{r}a_{s}\right\rangle _{[6]}T_{i}\left( y_{1}\right)
T_{k}\left( y_{2}\right) T_{m}\left( y_{3}\right) T_{n}\left( y_{4}\right)
T_{r}\left( y_{5}\right) T_{s}(y_{6}),  \label{C6}
\end{equation}%
where 
\begin{eqnarray}
\left\langle a_{i}a_{k}a_{m}a_{n}a_{r}a_{s}\right\rangle _{[6]} &\equiv
&\left\langle a_{i}a_{k}a_{m}a_{n}a_{r}a_{s}\right\rangle -[\underset{15%
\text{ variations}}{\underbrace{\left\langle a_{i}a_{k}\right\rangle
\left\langle a_{m}a_{n}a_{r}a_{s}\right\rangle +...}}]-  \notag \\
&&[\underset{10\text{ variations}}{\underbrace{\left\langle
a_{i}a_{k}a_{m}\right\rangle \left\langle a_{n}a_{r}a_{s}\right\rangle +...}}%
]+2[\underset{15\text{ variations}}{\underbrace{\left\langle
a_{i}a_{k}\right\rangle \left\langle a_{m}a_{n}\right\rangle \left\langle
a_{r}a_{s}\right\rangle +...}}].
\end{eqnarray}

The leading term is 
\begin{equation}
\langle a_{i}^{6}\rangle _{\lbrack 6]}\equiv \langle a_{i}^{6}\rangle
-15\langle a_{i}^{2}\rangle \langle a_{i}^{4}\rangle -10\langle
a_{i}^{3}\rangle ^{2}+30\langle a_{i}^{2}\rangle ^{3}.
\end{equation}

\section{Comments and conclusions}

We propose to measure higher order cumulants of event-by-event fluctuations
of rapidity distribution. The multi-particle rapidity distributions $\rho
_{n}(y_{1},\dots ,y_{n})$ can be expanded in the basis of orthogonal
polynomials $T_{i_{1}}(y_{1})\dots T_{i_{n}}(y_{n})$ with coefficients $%
\langle a_{i_{1}}\dots a_{i_{n}}\rangle $. The coefficients have
contributions from the rapidity shape fluctuations of the distribution
function, that we are after, and from the short-range correlations.
Calculating higher cumulants of such averages reduces the contribution from
short-range correlations. For example, in the fourth order cumulant $\langle
T_{i}(y_{a})T_{j}(y_{b})T_{k}(y_{c})T_{l}(y_{d})\rangle _{\lbrack 4]}$ the
contribution from two- and three-particle short-range correlations is
removed and, if the fourth-order short-range correlations are neglected, it
can be directly compared to the fourth cumulant of the expansion
coefficients $\langle a_{i}a_{j}a_{k}a_{l}\rangle _{\lbrack 4]}$.

The extracted value of the higher cumulants $\langle a_{i_{1}}\dots
a_{i_{n}}\rangle _{\lbrack n]}$ can be compared to predictions of models of
event-by-event fluctuations of rapidity distributions (\ref{ro-an}). The
cumulant of the coefficients $a_{i}$ can be calculated in models once their
event-by-event distribution is known. The proposed method to study the
cumulants of the expansion coefficients does not rely on the precise model
assumptions for fluctuations of rapidity distributions. It remains a subject
of further studies to calculate all significant four or six order cumulants,
in different models of energy deposition in hadronic collisions. The direct
calculation of higher order cumulants from experiment or Monte Carlo events
requires a very larger statistics. The limited statistics available has
prevented us from applying the method to realistic hydrodynamic events for
Pb-Pb collisions at the LHC.

Finally, we would like to emphasize that our method is applicable not only
to symmetric A+A collisions but also to asymmetric interactions including
p+A. It would be also interesting to perform measurement in p+p collisions,
where the internal quark (diquark \cite{abab:qd}) structure of a proton
should results in, e.g., nonzero asymmetric term $\langle a_{1}^{n}\rangle
_{[n]}$, $n=2,4,6$.

\vspace{\baselineskip} \noindent \textbf{Acknowledgments:} \newline
{}\newline
Research supported by the Ministry of Science and Higher Education (MNiSW), by founding from the Foundation for Polish Science, and by the National Science Centre (Narodowe Centrum Nauki), Grant Nos. DEC-2014/15/B/ST2/00175, DEC-2012/05/B/ST2/02528, and in part by DEC-2013/09/B/ST2/00497.


\begin{thebibliography}{99}

%\cite{ATLAS:2015kla}
\bibitem{ATLAS:2015kla} 
  The ATLAS collaboration [ATLAS Collaboration],
  %``Measurement of two-particle pseudorapidity correlations in lead-lead collisions at $\sqrt{s_{NN}}$ = 2.76 TeV with the ATLAS detector,''
  ATLAS-CONF-2015-020; http://cds.cern.ch/record/2029370
  %%CITATION = ATLAS-CONF-2015-020;%%

%\cite{Bialas:2011bz}
\bibitem{Bialas:2011bz} 
  A.~Bialas, A.~Bzdak and K.~Zalewski,
  %``Hidden Asymmetry and Long Range Rapidity Correlations,''
  Phys.\ Lett.\ B {\bf 710}, 332 (2012)
  [arXiv:1107.1215 [hep-ph]].
  %%CITATION = ARXIV:1107.1215;%%
  
%\cite{Bzdak:2012tp}
\bibitem{Bzdak:2012tp} 
  A.~Bzdak and D.~Teaney,
  %``Longitudinal fluctuations of the fireball density in heavy-ion collisions,''
  Phys.\ Rev.\ C {\bf 87}, no. 2, 024906 (2013)
  [arXiv:1210.1965 [nucl-th]].
  %%CITATION = ARXIV:1210.1965;%%
  %12 citations counted in INSPIRE as of 04 sept. 2015
  
%\cite{Bozek:2010vz}
\bibitem{Bozek:2010vz} 
  P.~Bozek, W.~Broniowski and J.~Moreira,
  %``Torqued fireballs in relativistic heavy-ion collisions,''
  Phys.\ Rev.\ C {\bf 83}, 034911 (2011)
  [arXiv:1011.3354 [nucl-th]].
  %%CITATION = ARXIV:1011.3354;%%
  %38 citations counted in INSPIRE as of 04 sept. 2015
  
%\cite{Bozek:2015bha}
\bibitem{Bozek:2015bha} 
  P.~Bozek, W.~Broniowski and A.~Olszewski,
  %``Hydrodynamic modeling of pseudorapidity flow correlations in relativistic heavy-ion collisions and the torque effect,''
  Phys.\ Rev.\ C {\bf 91}, 054912 (2015)
  [arXiv:1503.07425 [nucl-th]].
  %%CITATION = ARXIV:1503.07425;%%
  %1 citations counted in INSPIRE as of 04 sept. 2015
  
%\cite{Bialas:2011xk}
\bibitem{Bialas:2011xk} 
  A.~Bialas and K.~Zalewski,
  %``Multibin long-range correlations,''
  Nucl.\ Phys.\ A {\bf 860}, 56 (2011)
  [arXiv:1101.1907 [hep-ph]].
  %%CITATION = ARXIV:1101.1907;%%
  
%\cite{Olszewski:2015xba}
\bibitem{Olszewski:2015xba} 
  A.~Olszewski and W.~Broniowski,
  %``Multibin correlations in a superposition approach to relativistic heavy-ion collisions,''
  Phys.\ Rev.\ C {\bf 92}, no. 2, 024913 (2015)
  [arXiv:1502.05215 [nucl-th]].
  %%CITATION = ARXIV:1502.05215;%%
  %1 citations counted in INSPIRE as of 04 sept. 2015

%\cite{Cheng:2011hz}
\bibitem{Cheng:2011hz} 
  Y.~Cheng, Y.~-L.~Yan, D.~-M.~Zhou, X.~Cai, B.~-H.~Sa and L.~P.~Csernai,
  %``Longitudinal Fluctuations in Partonic and Hadronic Initial State,''
  Phys.\ Rev.\ C {\bf 84}, 034911 (2011)
  [arXiv:1106.3371 [hep-ph]].
  %%CITATION = ARXIV:1106.3371;%%
  
%\cite{Csernai:2012mh}
\bibitem{Csernai:2012mh} 
  L.~P.~Csernai, G.~Eyyubova and V.~K.~Magas,
  %``New method for measuring longitudinal fuctuations and directed flow in ultrarelativistic heavy ion reactions,''
  Phys.\ Rev.\ C {\bf 86}, 024912 (2012)
  [arXiv:1204.5885 [hep-ph]].
  %%CITATION = ARXIV:1204.5885;%%

%\cite{Pang:2012he}
\bibitem{Pang:2012he} 
  L.~Pang, Q.~Wang and X.~-N.~Wang,
  %``Effects of initial flow velocity fluctuation in event-by-event (3+1)D hydrodynamics,''
  Phys.\ Rev.\ C {\bf 86}, 024911 (2012)
  [arXiv:1205.5019 [nucl-th]].
  %%CITATION = ARXIV:1205.5019;%%

%\cite{Vovchenko:2013viu}
\bibitem{Vovchenko:2013viu} 
  V.~Vovchenko, D.~Anchishkin and L.~P.~Csernai,
  %``Longitudinal fluctuations of the center of mass of the participants in heavy-ion collisions,''
  Phys.\ Rev.\ C {\bf 88}, no. 1, 014901 (2013)
  [arXiv:1306.5208 [nucl-th]].
  %%CITATION = ARXIV:1306.5208;%%
  %17 citations counted in INSPIRE as of 04 sept. 2015

%\cite{Jia:2014ysa}
\bibitem{Jia:2014ysa} 
  J.~Jia and P.~Huo,
  %``Forward-backward eccentricity and participant-plane angle fluctuations and their influences on longitudinal dynamics of collective flow,''
  Phys.\ Rev.\ C {\bf 90}, no. 3, 034915 (2014)
  [arXiv:1403.6077 [nucl-th]].
  %%CITATION = ARXIV:1403.6077;%%
  %16 citations counted in INSPIRE as of 04 sept. 2015
  
%\cite{Pang:2014pxa}
\bibitem{Pang:2014pxa} 
  L.~G.~Pang, G.~Y.~Qin, V.~Roy, X.~N.~Wang and G.~L.~Ma,
  %``Longitudinal decorrelation of anisotropic flows in heavy-ion collisions at the CERN Large Hadron Collider,''
  Phys.\ Rev.\ C {\bf 91}, no. 4, 044904 (2015)
  [arXiv:1410.8690 [nucl-th]].
  %%CITATION = ARXIV:1410.8690;%%
  %5 citations counted in INSPIRE as of 04 sept. 2015  

%\cite{Voloshin:2008dg}
\bibitem{Voloshin:2008dg} 
  S.~A.~Voloshin, A.~M.~Poskanzer and R.~Snellings,
  %``Collective phenomena in non-central nuclear collisions,''
  arXiv:0809.2949 [nucl-ex].
  %%CITATION = ARXIV:0809.2949;%%
  
%\cite{Alver:2010gr}
\bibitem{Alver:2010gr} 
  B.~Alver and G.~Roland,
  %``Collision geometry fluctuations and triangular flow in heavy-ion collisions,''
  Phys.\ Rev.\ C {\bf 81}, 054905 (2010)
  [Erratum-ibid.\ C {\bf 82}, 039903 (2010)]
  [arXiv:1003.0194 [nucl-th]].
  %%CITATION = ARXIV:1003.0194;%%
  
%\cite{Bialas:1976ed,Bialas:2004su}
%\cite{Bialas:1976ed}
\bibitem{Bialas:1976ed} 
  A.~Bialas, M.~Bleszynski and W.~Czyz,
  %``Multiplicity Distributions in Nucleus-Nucleus Collisions at High-Energies,''
  Nucl.\ Phys.\ B {\bf 111}, 461 (1976).
  %%CITATION = NUPHA,B111,461;%%

%\cite{Bialas:2004su}
\bibitem{Bialas:2004su} 
  A.~Bialas and W.~Czyz,
  %``Wounded nucleon model and Deuteron-Gold collisions at RHIC,''
  Acta Phys.\ Polon.\ B {\bf 36}, 905 (2005)
  [hep-ph/0410265].
  %%CITATION = HEP-PH/0410265;%%
  
%\cite{Borghini:2000sa}
\bibitem{Borghini:2000sa} 
  N.~Borghini, P.~M.~Dinh and J.~Y.~Ollitrault,
  %``A New method for measuring azimuthal distributions in nucleus-nucleus collisions,''
  Phys.\ Rev.\ C {\bf 63}, 054906 (2001)
  [nucl-th/0007063].
  %%CITATION = NUCL-TH/0007063;%%
  %170 citations counted in INSPIRE as of 06 sept. 2015
  
%\cite{Borghini:2001vi}
\bibitem{Borghini:2001vi} 
  N.~Borghini, P.~M.~Dinh and J.~Y.~Ollitrault,
  %``Flow analysis from multiparticle azimuthal correlations,''
  Phys.\ Rev.\ C {\bf 64}, 054901 (2001)
  [nucl-th/0105040].
  %%CITATION = NUCL-TH/0105040;%%
  %275 citations counted in INSPIRE as of 06 sept. 2015
  
%\cite{Khachatryan:2015waa}
\bibitem{Khachatryan:2015waa} 
  V.~Khachatryan {\it et al.} [CMS Collaboration],
  %``Evidence for Collective Multiparticle Correlations in p-Pb Collisions,''
  Phys.\ Rev.\ Lett.\  {\bf 115}, no. 1, 012301 (2015)
  [arXiv:1502.05382 [nucl-ex]].
  %%CITATION = ARXIV:1502.05382;%%
  %9 citations counted in INSPIRE as of 06 sept. 2015
  
%\cite{Jia:2015jga}
\bibitem{Jia:2015jga} 
  J.~Jia, S.~Radhakrishnan and M.~Zhou,
  %``Forward-backward multiplicity fluctuation and longitudinal harmonics in high-energy nuclear collisions,''
  arXiv:1506.03496 [nucl-th].
  %%CITATION = ARXIV:1506.03496;%%
  %2 citations counted in INSPIRE as of 04 sept. 2015
  
%\cite{Bilandzic:2010jr}
\bibitem{Bilandzic:2010jr} 
  A.~Bilandzic, R.~Snellings and S.~Voloshin,
  %``Flow analysis with cumulants: Direct calculations,''
  Phys.\ Rev.\ C {\bf 83}, 044913 (2011)
  [arXiv:1010.0233 [nucl-ex]].
  %%CITATION = ARXIV:1010.0233;%%
  %115 citations counted in INSPIRE as of 04 sept. 2015
   
%\cite{abab:qd}
\bibitem{abab:qd} 
  A.~Bialas and A.~Bzdak,
  Phys.\ Rev.\ C {\bf 77}, 034908 (2008)
  [arXiv:0707.3720 [hep-ph]]; Phys.\ Lett.\ B {\bf 649}, 263 (2007)
  [nucl-th/0611021].
  %%CITATION = ARXIV:0707.3720;%%

\end{thebibliography}
\end{document}